\pdfoutput=1

\documentclass[journal]{IEEEtran}
\usepackage[caption=false,font=footnotesize]{subfig}
\usepackage{graphicx}
\begin{document}
\title{Imaging Pion Showers with the CALICE\\
  Analogue Hadron Calorimeter}

\author{Nils~Feege on behalf of the CALICE collaboration%
  \thanks{Nils~Feege is with the Institute for Experimental Physics, University of Hamburg, Hamburg, Germany and DESY, Hamburg, Germany (e-mail: nils.feege@desy.de).}%
}

\maketitle
\thispagestyle{empty}

\begin{abstract}

  The CALICE collaboration investigates different technology options for highly granular calorimeters for detectors at a future electron-positron collider. One of the devices constructed and tested by the collaboration is a 1\,m$\bf^{3}$ prototype for an imaging scintillator-steel sampling calorimeter for hadrons with analogue readout (AHCAL). The light from 7608 small scintillator cells is detected with silicon photomultipliers. The AHCAL has been successfully operated during electron and hadron test-beam measurements at DESY, CERN, and Fermilab since 2005. The collected data allow for evaluating the novel technologies employed. In addition, these data provide a valuable basis for validating pion cascade simulations. This paper presents the current status of comparisons between the AHCAL data and predictions from different Monte Carlo models implemented in \textbf{\textsc{Geant4}}. The comparisons cover the total visible energy, longitudinal and radial shower profiles, and the shower substructure. Furthermore, this paper discusses a software compensation algorithm for improving the energy resolution of the AHCAL for single pions.

\end{abstract}


\section{Introduction}
\IEEEPARstart{T}{he} CALICE collaboration investigates new imaging calorimeter technologies for precision measurements at a future electron-positron linear collider. These calorimeters feature a fine granularity in both longitudinal and transverse direction. The high spacial resolution is needed to fulfil the shower separation requirement of Particle Flow reconstruction algorithms. The Particle Flow approach aims for a jet energy resolution of 3-4\% at the proposed International Linear Collider \cite{snowmass1}, \cite{pflow1}, \cite{pflow2}.

\begin{figure}[!t]
  \centering
  \includegraphics[width=2.5in]{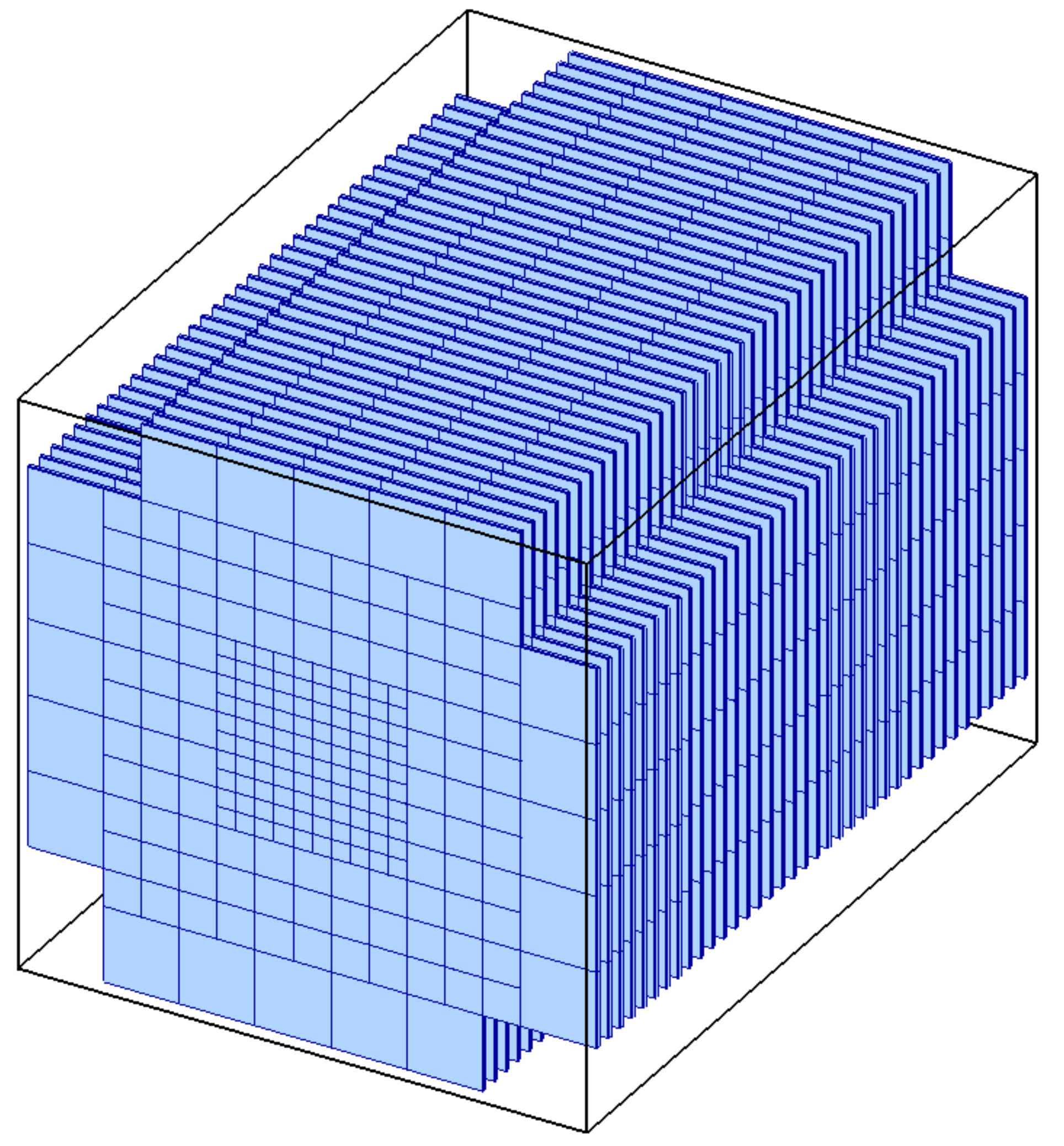}
  \caption{The 38 scintillator layers of the AHCAL prototype. Each layer is 5\,mm thick and is pieced together from 216 tiles measuring $3 \times 3\,\mathrm{cm}^{2}$, $6 \times 6\,\mathrm{cm}^{2}$, or $12 \times 12\,\mathrm{cm}^{2}$. }
  \label{fig_ahcal}
\end{figure}

\begin{figure}[!t]
  \centering
  \includegraphics[width=2.5in]{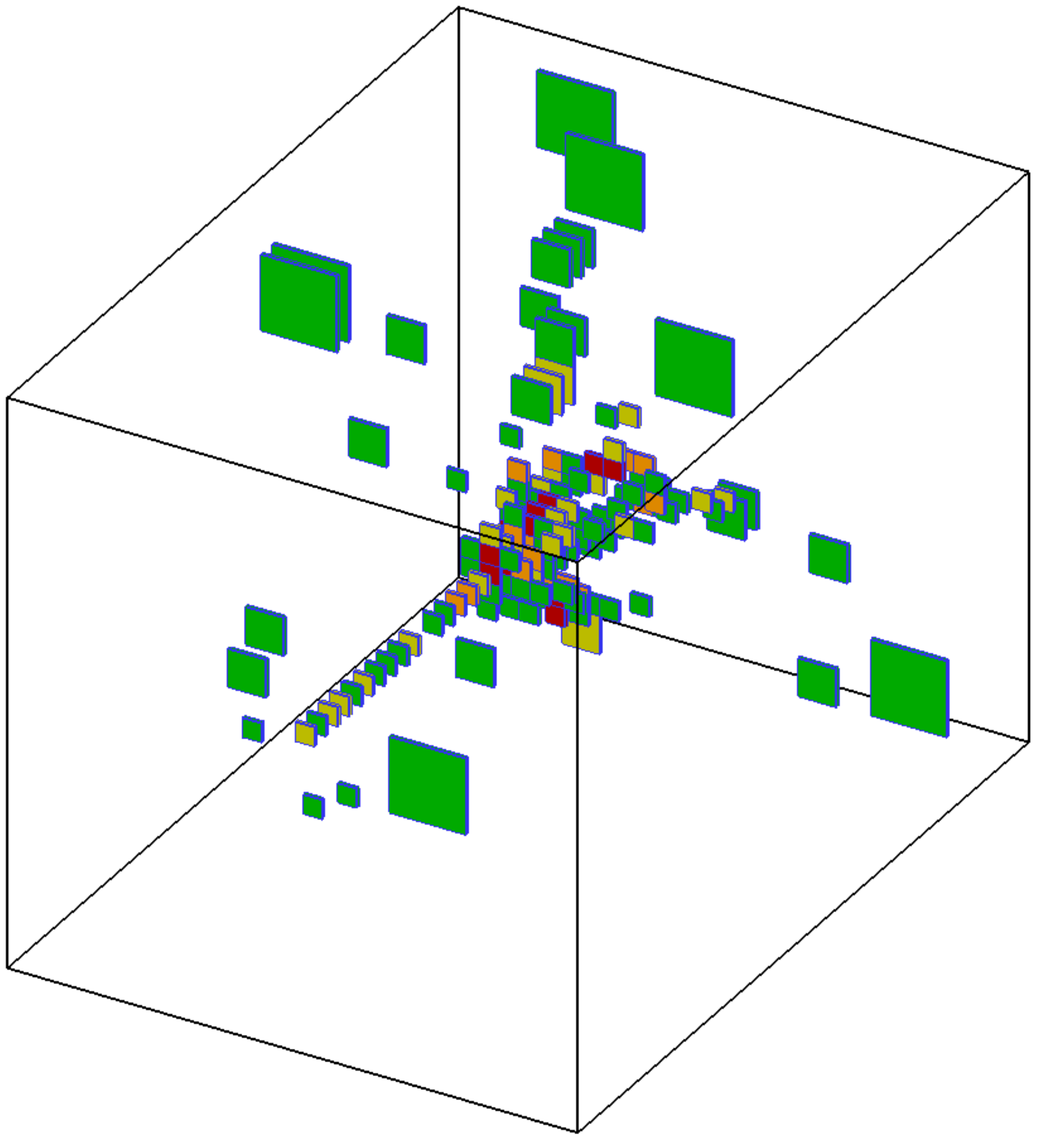}
  \caption{Measurement of the energy deposited by a 10\,GeV negative pion in the AHCAL. The primary ionisation track, the location of the first inelastic pion-nucleus interaction and the development of the hadronic cascade are visible. }
  \label{fig_10GeVevent}
\end{figure}

CALICE has designed and constructed several prototypes for electromagnetic and hadron calorimeters. One of these devices is the prototype for an analogue hadron calorimeter (AHCAL). This prototype is a 1\,$\mathrm{m}^3$ scintillator-steel sampling calorimeter with 38 sensitive layers. The steel plates are 2\,cm thick. Each scintillator layer is pieced together from separate tiles. The tiles are 5\,mm thick and their lateral size varies between $3 \times 3\,\mathrm{cm}^{2}$, $6 \times 6\,\mathrm{cm}^{2}$, and $12 \times 12\,\mathrm{cm}^{2}$. Figure \ref{fig_ahcal} illustrates the layout of the sensitive layers. A wavelength-shifting fibre collects the light from each tile and guides it to a silicon photomultiplier (SIPM, c.f. \cite{sipm1}). The SIPM is directly attached to the tile. With a total of 7608 readout channels, the AHCAL prototype represents the first large-scale application of SIPMs. The detector is described in more detail in \cite{ahcal1}. Figure \ref{fig_10GeVevent} presents a cascade from a 10\,GeV pion measured in the AHCAL during test-beam operation. This measurement illustrates the imaging capabilities of the device.

\begin{figure}[!t]
  \centering
  \includegraphics[width=3.0in]{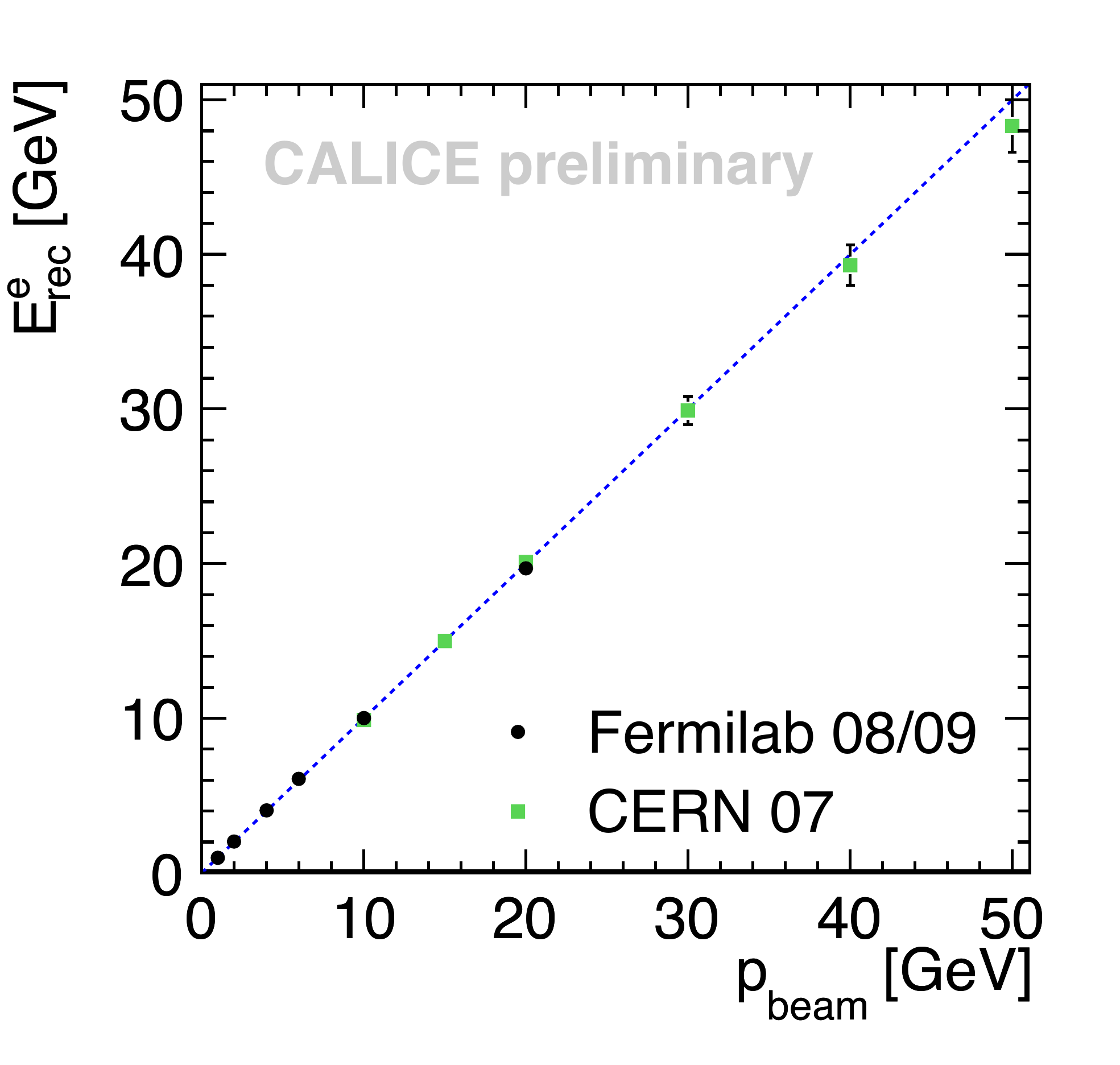}
  \caption{ Mean reconstructed electron (Fermilab) and positron (CERN) energies at different beam momenta. The data confirm a linear response (indicated by the dashed line) up to 30\,GeV. The non-linearity above 30\,GeV is of the same order as the calibration uncertainty (3\% at 50 GeV).}
  \label{fig_eLin}
\end{figure}

The response of the calorimeter cells is equalised and calibrated using broad beams of muons acting as minimum ionising particles, i.e. the energy measured in each cell is expressed in multiples of the energy deposited by a muon in this cell (MIP). To suppress noise, only cells with a deposited energy of at least 0.5\,MIP are considered for analysis. After correcting for the SIPM non-linearity, the uncertainty on the calibrated signal is roughly 3\%. The electromagnetic scale, i.e. the conversion factor from MIP to GeV for an electromagnetic cascade, is determined using electron or positron data from test-beam measurements. Data sets covering different energy ranges (energy and momentum are used interchangeably throughout this paper) and acquired under different operation conditions (CERN 2007: 10-50 GeV positrons; Fermilab 2009: 1-20\,GeV electrons) yield conversion factors that agree within their uncertainties. In addition, these data confirm the linearity of the electron (and positron) response up to 30\,GeV. Figure~\ref{fig_eLin} shows the AHCAL response to electrons and positrons of different beam momenta. The observed non-linearity above 30\,GeV (up to 3\% at 50\,GeV) is of the same order as the calibration uncertainty and is attributed to remaining effects from the SIPM saturation. More on the performance of the calorimeter can be found in \cite{ahcal2} and \cite{can34}.

A detailed model of the AHCAL has been implemented in Mokka \cite{mokka} (version 7.02). Mokka is a \textsc{Geant4} \cite{geant4} based Monte Carlo application and is capable of simulating test-beam setups and full detector geometries. The agreement between the MIP/GeV factor and other observables extracted from electron simulations and test-beam data validates the simulation of the various detector characteristics \cite{ahcal2}.

This paper presents comparisons between data (collected with the AHCAL at CERN in 2007) and \textsc{Geant4} simulations for negative pions between 8\,GeV and 80\,GeV beam momentum. Several \textsc{Geant4} physics lists for the simulation of hadron interactions with matter exist. A physics list combines different models that are valid for different energy ranges. At the transition between models, one model is randomly chosen for each incoming particle. This paper concentrates on four physics lists: QGSP\_BERT, FTFP\_BERT, FTF\_BIC (\textsc{Geant4}, version 9.3) and CHIPS (\textsc{Geant4}, version 9.3.p01). Validation studies performed by LHC experiments favour the QGSP\_BERT physics list. Both FTFP\_BERT and FTF\_BIC have been re-tuned recently and are promising alternatives for QGSP\_BERT. The CHIPS physics list is still at an experimental stage. However, this list is particularly interesting because it uses only one model for all hadron energies. Table \ref{table_physlists} summarises the models used by these physics lists for simulating pions of different energies. A more detailed description of the physics lists and the individual models is given in \cite{eudetmemo}.

\begin{table}[!t]
  \renewcommand{\arraystretch}{1.5}
  \caption{Composition of \textsc{Geant4} physics lists.}
  \label{table_physlists}
  \centering
  \begin{tabular}{| l | | l | l |}
    \hline
    Physics list & Model (for $\pi^{\pm}$) & Energy range \\
    \hline
    QGSP\_BERT & Bertini cascade (BERT) & $\leq$ 9.9\,GeV \\
    & Low-energy parametrisation (LEP) & 9.5\,GeV - 25\,GeV \\
    & Quark-gluon string model (QGSP) & $\geq$ 12\,GeV  \\
    \hline
    FTFP\_BERT & Bertini cascade (BERT) & $\leq$ 5\,GeV \\
    & Fritiof string model (FTFP) & $\geq$ 4\,GeV  \\
    \hline
    FTF\_BIC        & Binary cascade (BIC) & $\leq$ 5\,GeV \\
    & Fritiof string model (FTF) & $\geq$ 4\,GeV  \\
    \hline
    CHIPS & Chiral-invariant phase space model & $\geq$ 0\,GeV \\
    \hline
  \end{tabular}
\end{table}

\section{Calorimeter Response to Pions}
\begin{figure}[!t]
  \centering
  \includegraphics[width=3.0in]{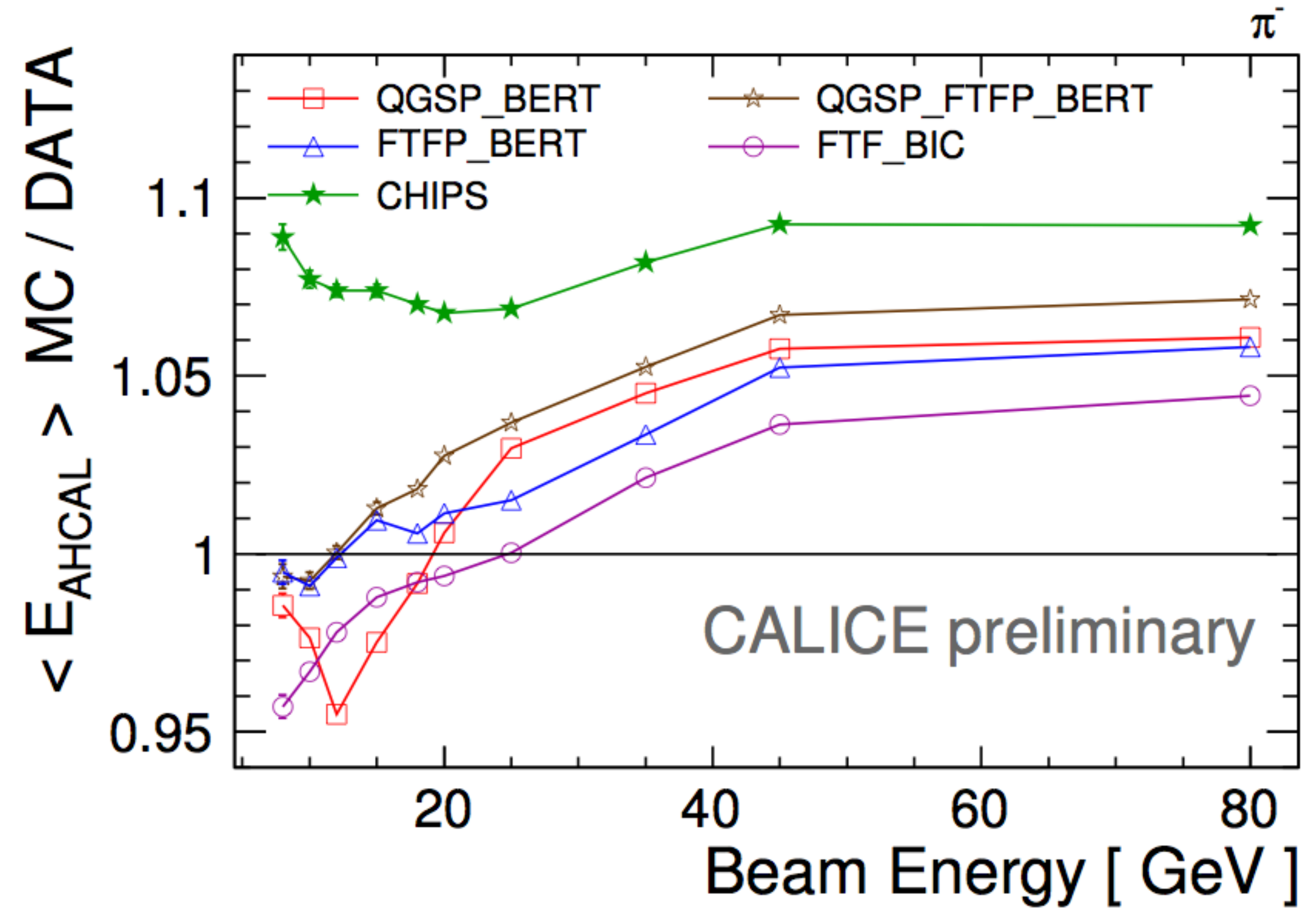}
  \caption{Ratio between the mean visible energy in the AHCAL from simulations and from data for negative pions at different beam energies. For all physics lists except for CHIPS, this ratio depends on the pion energy.}
  \label{fig_esum_ratio}
\end{figure}

Figure \ref{fig_esum_ratio} presents the ratio between the mean visible energy for negative pions from simulation and measurements for beam energies between 8\,GeV and 80\,GeV \cite{eudetmemo}. The predictions made by the CHIPS physics list show an energy independent overestimation of roughly 8\%. This overestimation is expected, because the low energy neutron cross-sections are not yet properly implemented in CHIPS. For the composite physics lists, the agreement between predictions and data depends on the particle energy. Both the QGSP\_BERT and FTFP\_BERT physics lists agree with data within 2\% at 8\,GeV and overestimate the detector response by 6\% at 80\,GeV. Between 8\,GeV and 25\,GeV, QGSP\_BERT shows a dip, which is attributed to the use of the LEP model as stopgap in this energy range. The FTF\_BIC physics list underestimates the visible energy by 4\% at 8\,GeV (Binary cascade and FTF) and gives 4\% too high energy deposition at 80\,GeV (FTF dominates the description of the first hard interaction).

\section{Topology of Pion Cascades}

\begin{figure}[!t]
  \centering
  \includegraphics[width=3.5in]{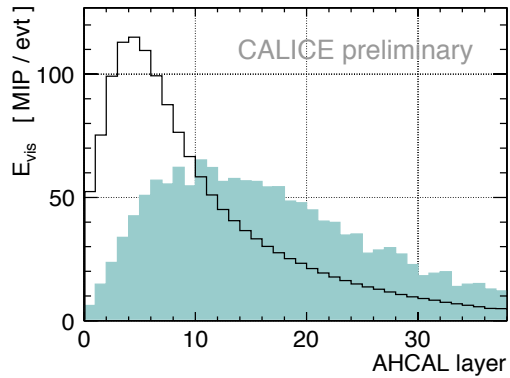}
  \caption{Longitudinal shower profile for negative pions (45\,GeV) relative to the calorimeter front (filled histogram) and relative to the first inelastic interaction (open histogram).}
  \label{fig_primint}
\end{figure}

Exploiting the high granularity of the AHCAL allows for determining the position of the first hard interaction of hadrons in the calorimeter \cite{can26}. According to Monte Carlo studies, the uncertainty of the algorithm applied for this determination is $\pm 1$ calorimeter layer (1 layer $\approx$ 3\,cm) in about 74\% of the events. The information on the position of the first interaction is used to investigate the average longitudinal shower profile relative to this point. Figure \ref{fig_primint} presents a longitudinal shower profile relative to the calorimeter front face (filled histogram) and a longitudinal shower profile realtive to the first interaction point (open histogram). The latter profile looks shorter and much more similar to an electromagnetic shower profile because the fluctuations of the shower starting position are removed.

\begin{figure*}[!t]
  \centerline{
    \subfloat[]{ \includegraphics[width=0.45\textwidth]{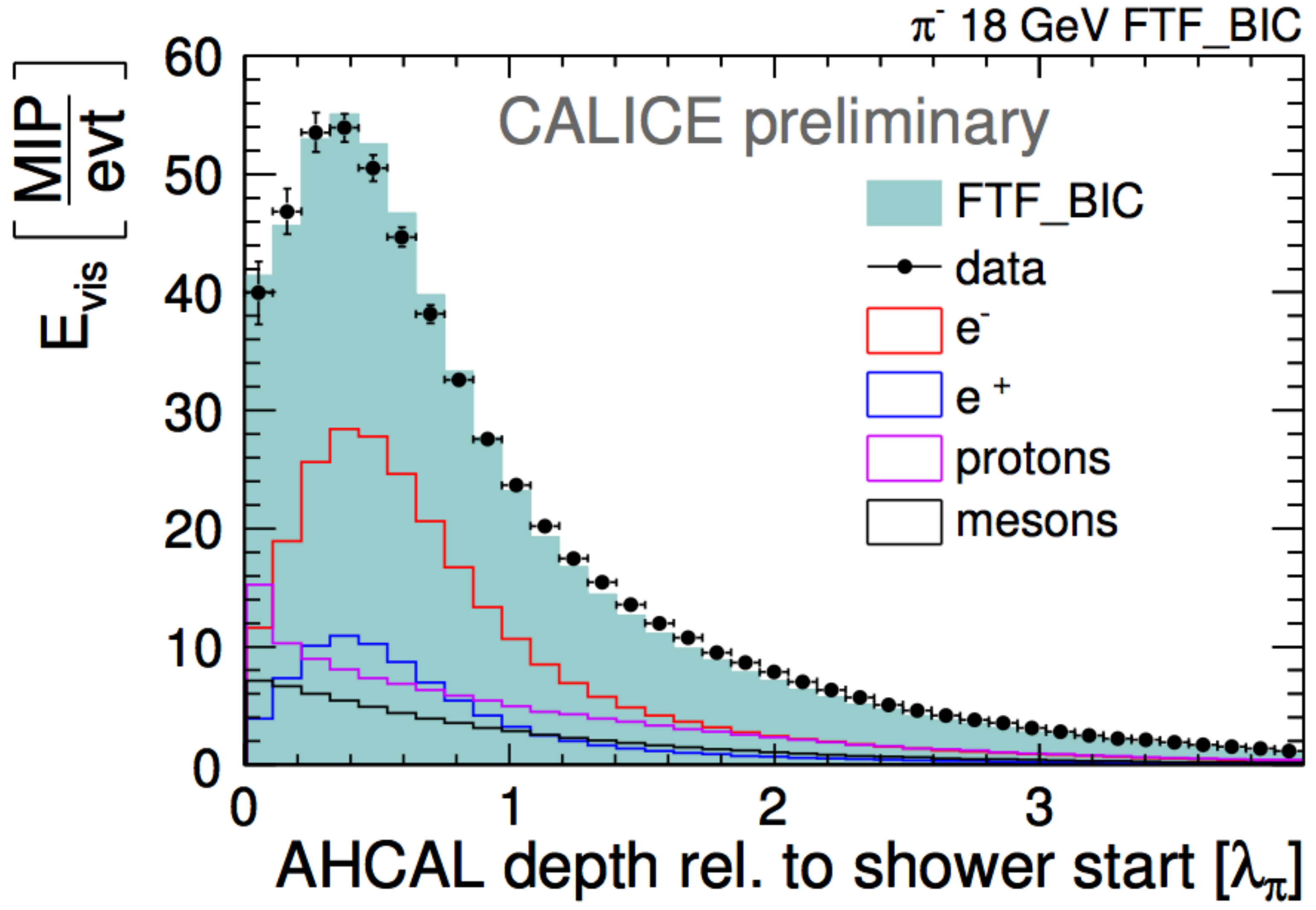}%
      \label{fig_lprof}  }
    \hfill
    \subfloat[]{\includegraphics[width=0.45\textwidth]{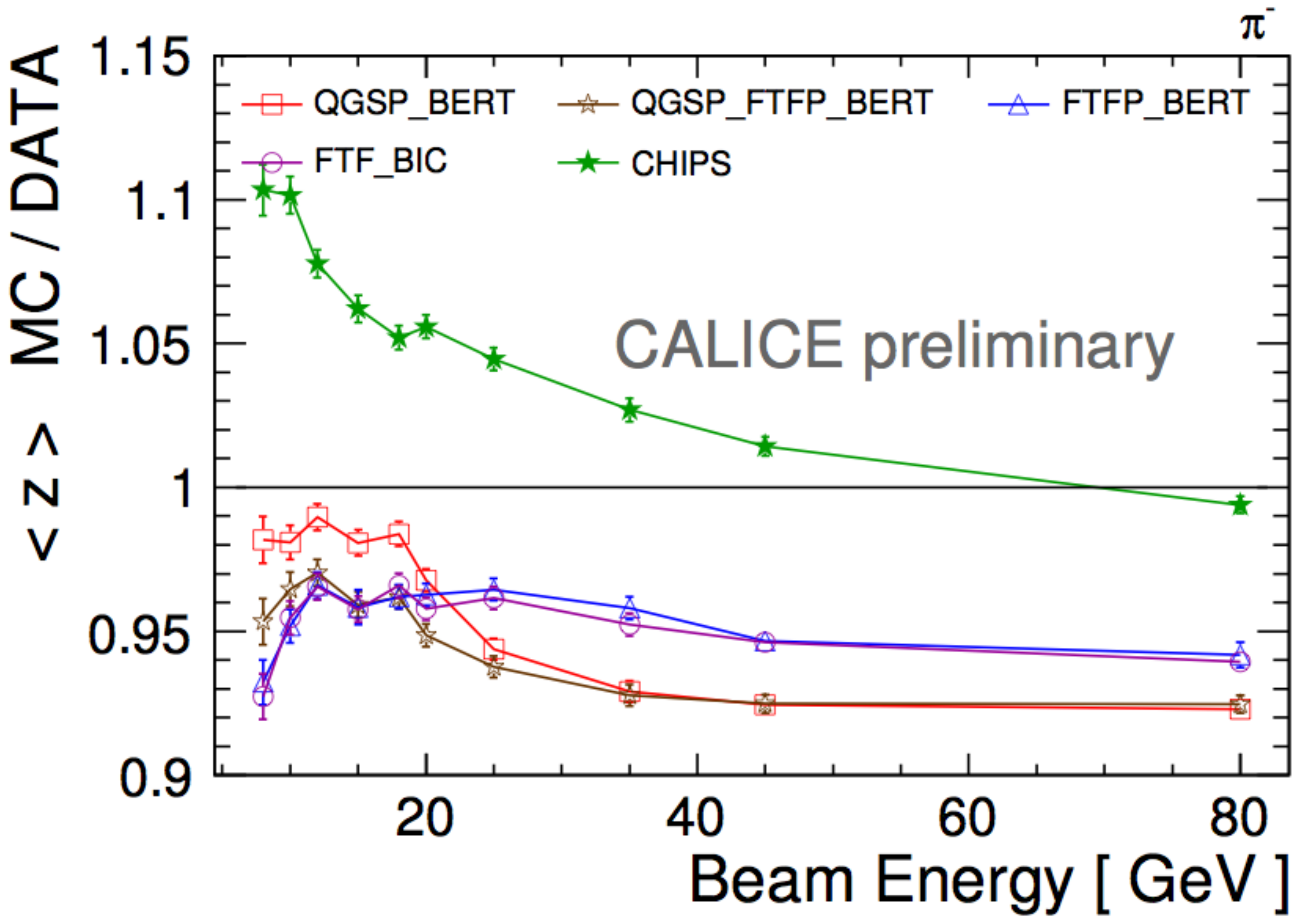}%
      \label{fig_lprof_ratio} } }
  \caption{(a) Longitudinal shower profile relative to the first interaction for negative pions (18\,GeV) in the AHCAL for data (black points) and simulation (FTF\_BIC physics list, filled histogram). The error bars include only the statistical uncertainty and the uncertainty introduced by the determination of the first inelastic scattering. The breakdown of the energy contribution from various particles in the shower (electrons, positrons, protons, and mesons) is shown. (b) Ratio between the mean of the longitudinal shower profiles (centre of gravity along beam axis) from simulation and data. CHIPS overestimates the shower depth, while the other physics lists behave opposite.}
\end{figure*}

Figure \ref{fig_lprof} shows the longitudinal shower profile relative to the position of the first hard interaction for negative 18\,GeV pions in the AHCAL \cite{eudetmemo}. Data are displayed as black points on top of the filled histogram from simulation (FTF\_BIC). The general shapes of these profiles agree. In addition to the profiles, Fig. \ref{fig_lprof} presents the breakdown of the energy contributions from various particles to the shower (electrons, positrons, protons, and mesons). This additional information helps to discuss which physics process contributes most in which phase of the shower development. The overall longitudinal shower shape is dominated by electromagnetic processes (electron and positron contribution to the entire profile). The hadrons only contribute significantly in the very first layers after the first hard interaction. Figure \ref{fig_lprof_ratio} shows the ratio between the mean of the longitudinal shower profiles (mean shower depth, centre of gravity along beam axis) from simulation and data. CHIPS overestimates the shower depth at energies below 80\,GeV. The other physics lists underestimate the shower depth for all energies presented.

\begin{figure*}[!t]
  \centerline{
    \subfloat[]{ \includegraphics[width=0.45\textwidth]{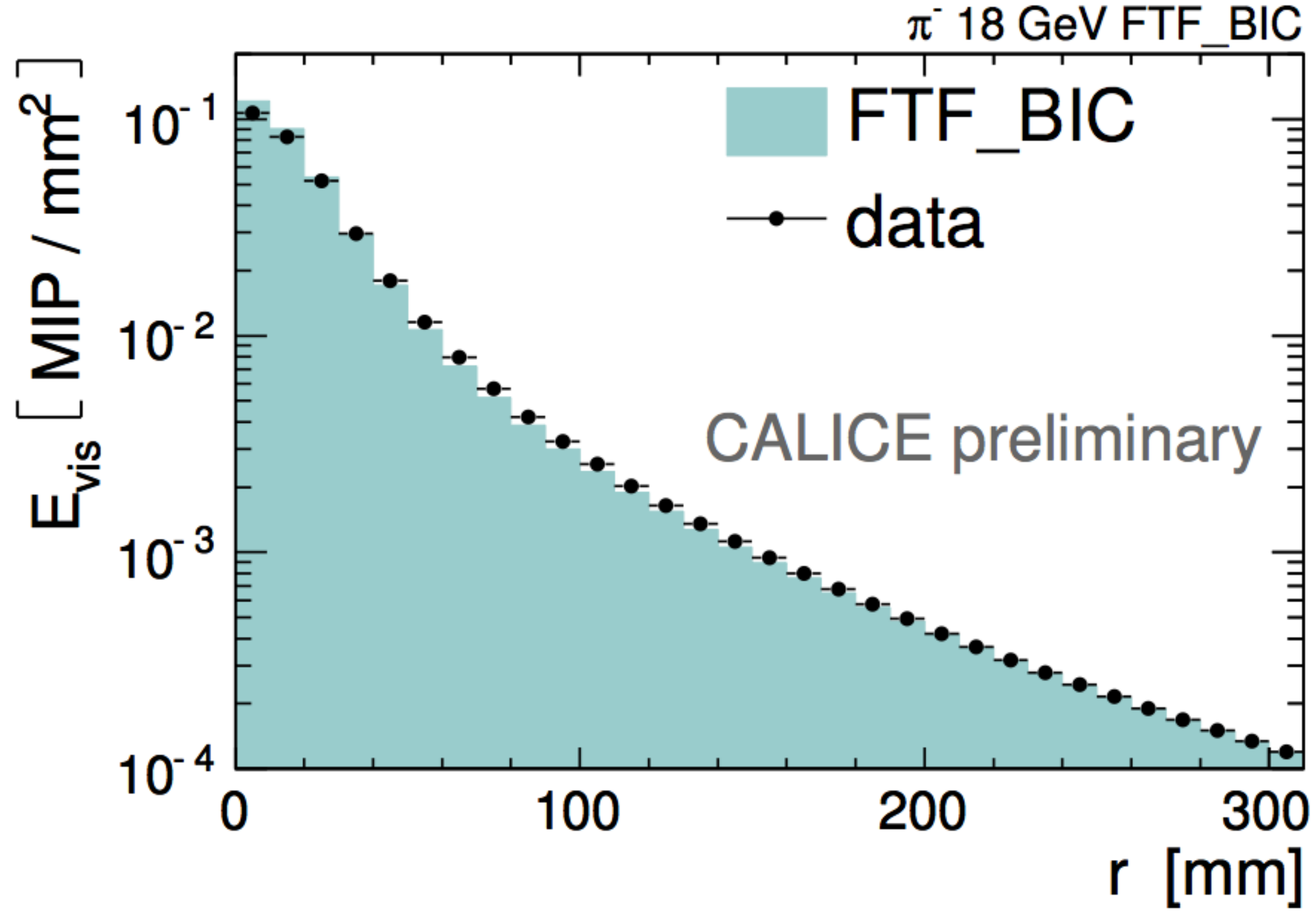}%
      \label{fig_rprof}  }
    \hfill
    \subfloat[]{\includegraphics[width=0.45\textwidth]{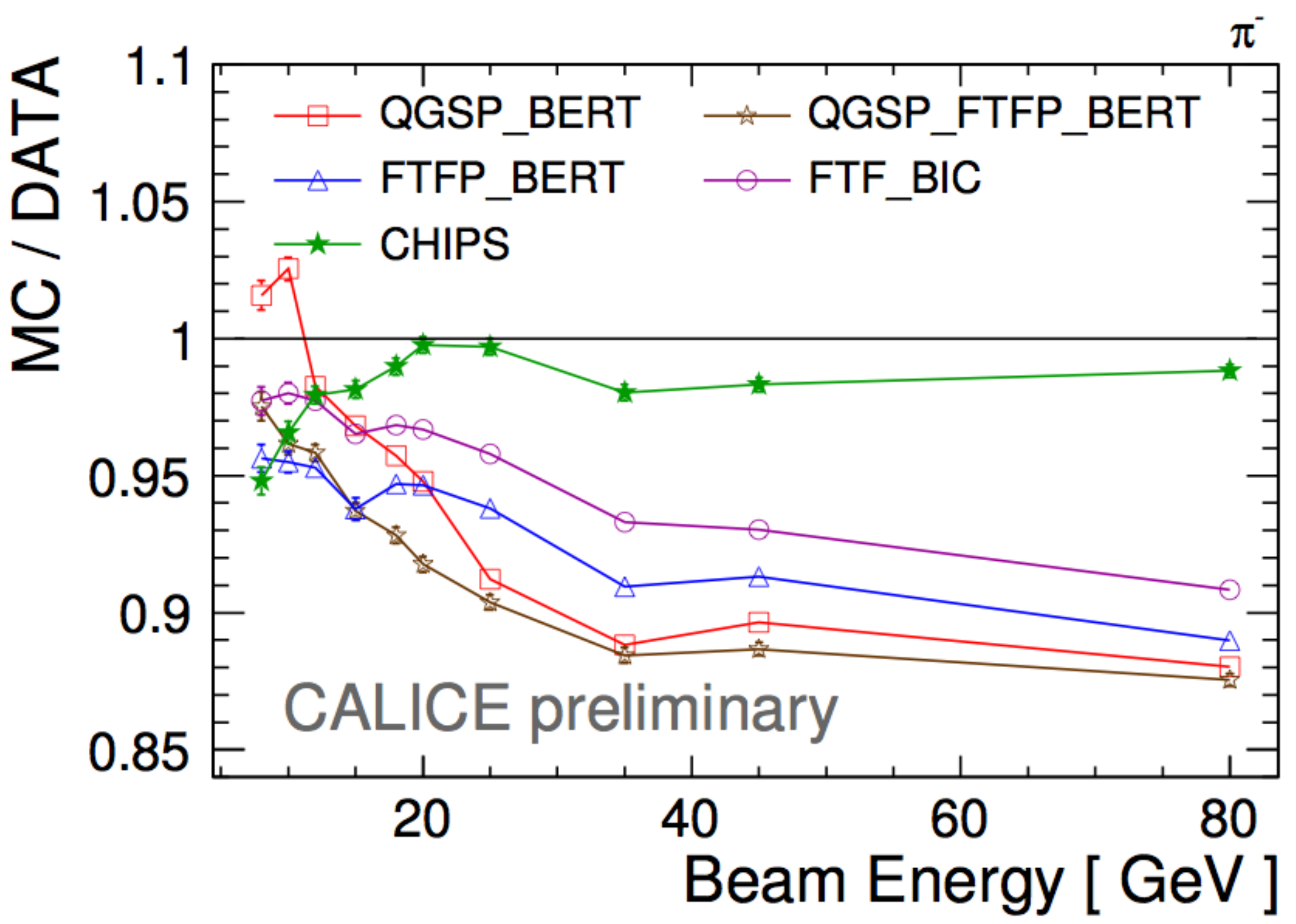}%
      \label{fig_rprof_ratio} } }
  \caption{(a) Radial shower profile for negative pions (18\,GeV) in the AHCAL for data (black points) and simulation (FTF\_BIC physics list, filled histogram). (b) Ratio between the mean of the radial shower profiles (mean shower radius) from simulation and data. CHIPS gives the best prediction of the shower radius.}
\end{figure*}

Figure \ref{fig_rprof} presents the radial shower profile for negative pions (18\,GeV) in the AHCAL for data (black points) and simulation (FTF\_BIC physics list, filled histogram). The overall shapes of the profiles agree. Figure \ref{fig_rprof_ratio} shows the ratio between the mean of the radial shower profiles (mean shower radius) from simulation and data. Except for QGSP\_BERT below 10\,GeV, all physics lists predict narrower showers than observed in data. CHIPS gives the best prediction of the shower radius.

\section{Track Multiplicity}
The high granularity of the CALICE AHCAL provides the capability for identifying track segments from secondary hadrons produced within hadron showers \cite{can22}. A simple tracking algorithm allows for finding tracks created by minimum ionising particles in the cascade. The algorithm relies on isolated hits and works on a layer-by-layer basis. The maximum angle $\theta$ between the beam direction and the reconstructed tracks is intrinsically limited by the algorithm. For tracks in the $3 \times 3\,\mathrm{cm}^{2}$ tiles, the limit is $\theta_{3\times3} \le 58^{\circ}$. For the larger tiles, the limit is correspondingly $\theta_{6\times6} \le 72^{\circ}$ and $\theta_{12\times12} \le 81^{\circ}$. The fake track rate of this algorithm is on the level of few per mille.

\begin{figure}[!H]
  \centering
  \includegraphics[width=3.5in]{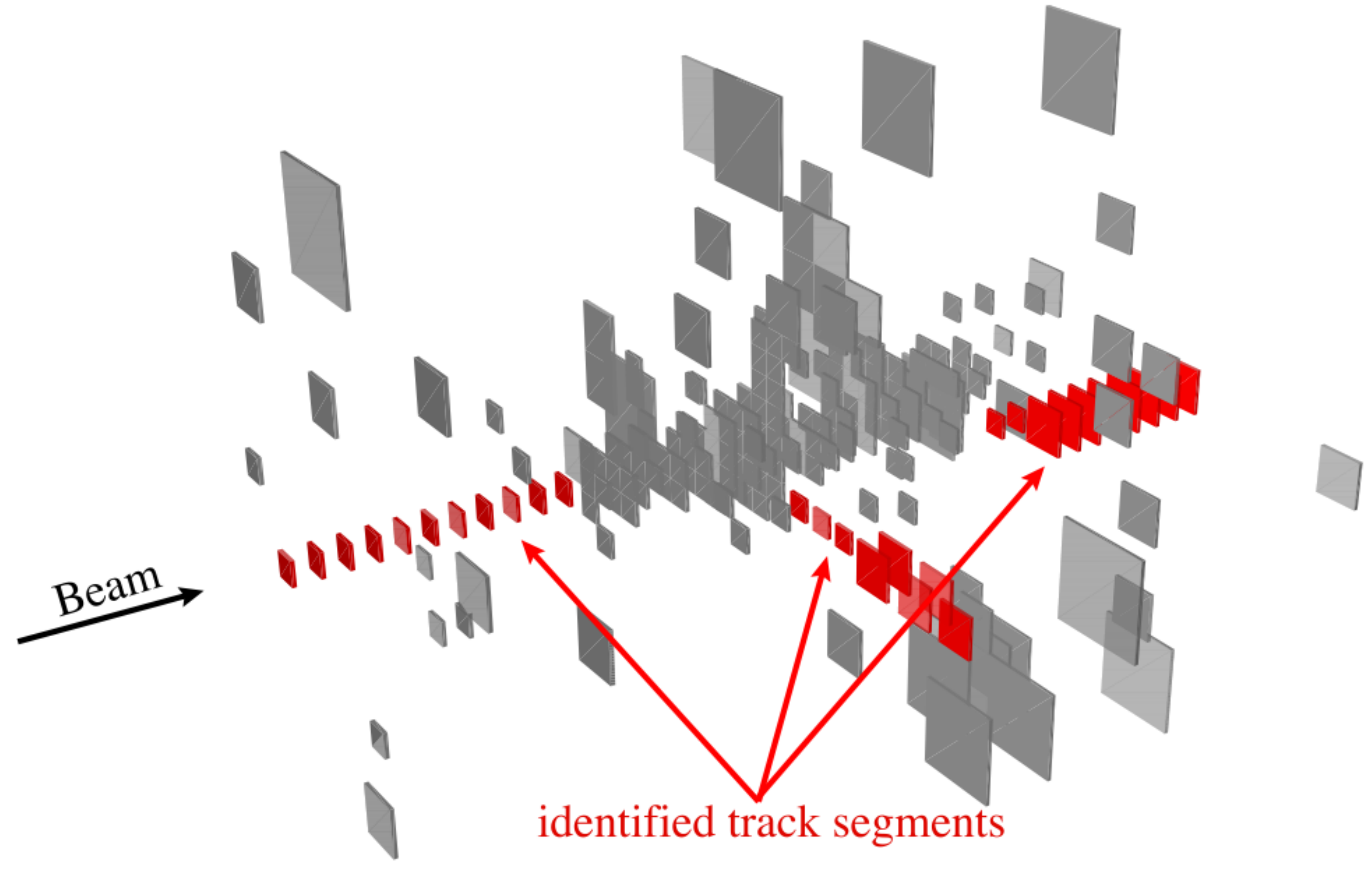}
  \caption{Energy deposited by a 20\,GeV pion shower in the AHCAL. The signals identified as track segments are highlighted in red.}
  \label{fig_evtTracks}
\end{figure}

\begin{figure}[!t]
  \centering
  \includegraphics[width=3.5in]{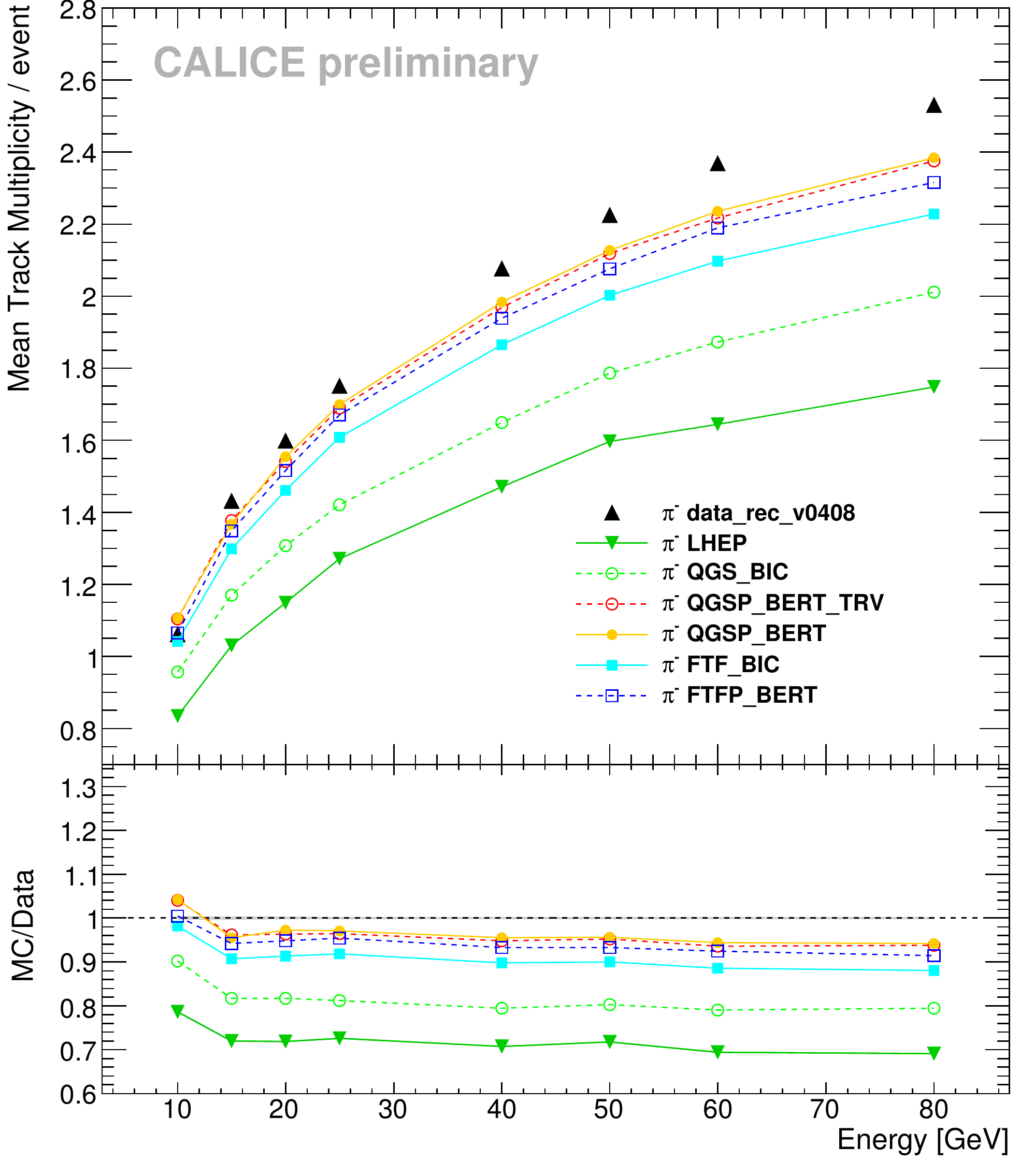}
  \caption{Average track multiplicity at different beam energies extracted from data and predicted by several physics lists.}
  \label{fig_trackMulti}
\end{figure}

Figure \ref{fig_evtTracks} shows a typical hadron cascade from a 20\,GeV negative pion in the AHCAL. Minimum ionising track segments of the incoming pion and of secondary particles are identified and highlighted in red. The track multiplicity is influenced by the shower topology and especially by the number of secondaries created. The average track multiplicity is shown as a function of the beam energy for data and various simulations in Fig. \ref{fig_trackMulti}. All physics lists predict too small track multiplicities. The physics list predicting values closest to data is QGSP\_BERT.

\section{Performance and Software Compensation}
Hadron cascades consist of an electromagnetic and a hadronic component. The electromagnetic part originates from neutral pions and eta mesons created in inelastic hadron-nucleus collisions. These mesons decay practically instantaneously into two photons, which give rise to electromagnetic cascades. Other hadrons involved in the cascade deposit part of their energy as invisible energy in the form of nuclear binding energy, nuclear recoil, or neutrinos. The AHCAL is a non-compensating calorimeter, i.e. the AHCAL response to the hadronic component of a pion cascade is smaller than the response to the electromagnetic component. The average electromagnetic fraction of a hadronic cascade increases with the momentum of the incident particle. The electromagnetic fraction and the invisible energy fluctuate strongly from one event to another. These two effects lead to a non-linear detector response to pions and degrade the detector resolution. Software compensation algorithms can be applied to correct for these effects.

Hadron showers with higher energy density (shower energy divided by shower volume) tend to have a larger reconstructed energy than showers with a smaller energy density for the same particle momentum. A clustering algorithm allows to select only energy depositions belonging to the actual hadron cascade and to exclude noise signals. Because of the high energy density in purely electromagnetic showers, a high energy density in a hadron shower is attributed to a large electromagnetic fraction, i.e. the cluster energy density gives an estimate of the electromagnetic fraction inside a cascade.

\begin{figure}[!t]
  \centering
  \includegraphics[width=3.5in]{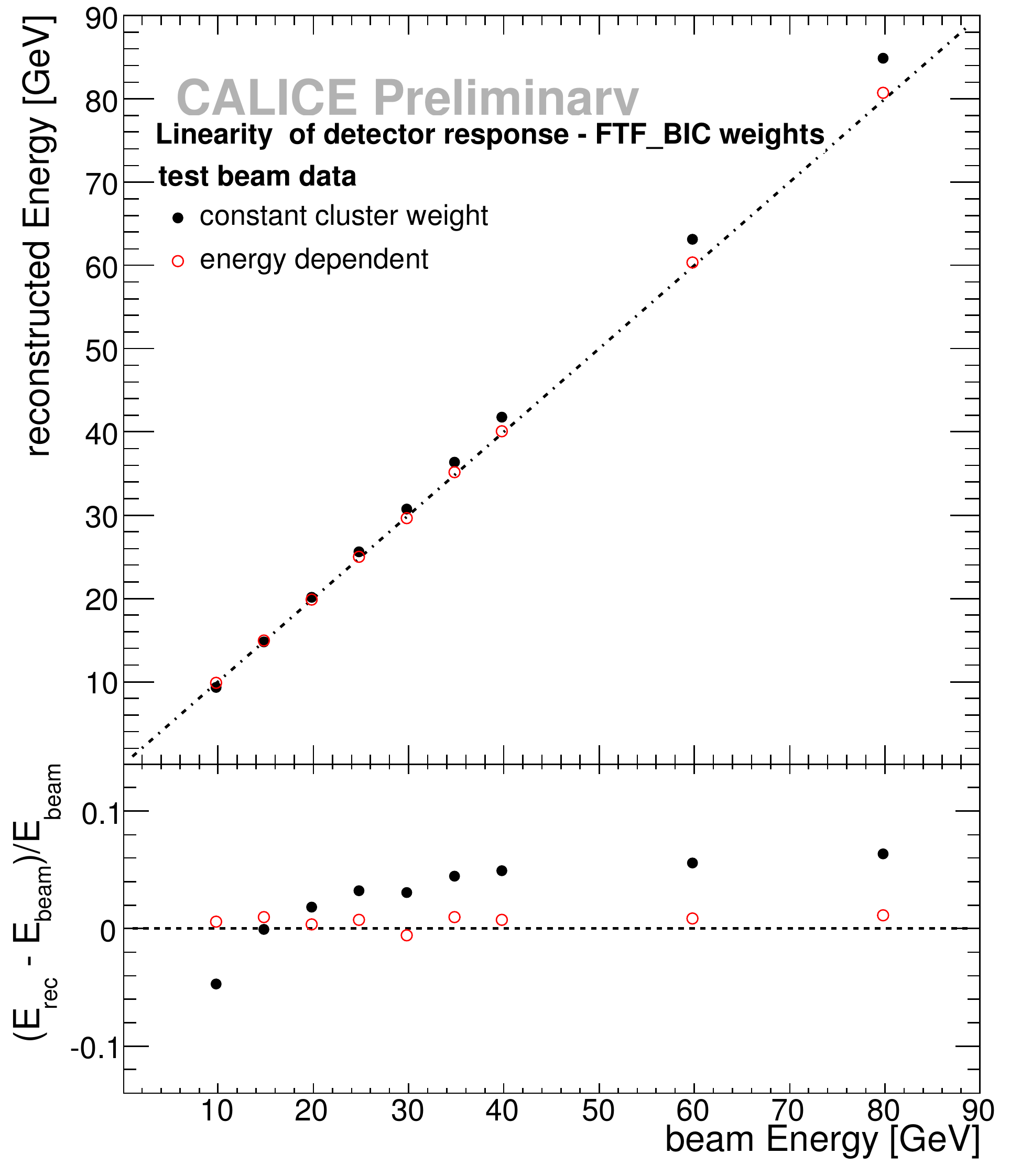}
  \caption{AHCAL linearity extracted from test-beam data with a single energy weight (filled circles) and energy weights based on the cluster energy density (open circles). }
  \label{fig_piLin}
\end{figure}

\begin{figure}[!t]
  \centering
  \includegraphics[width=3.5in]{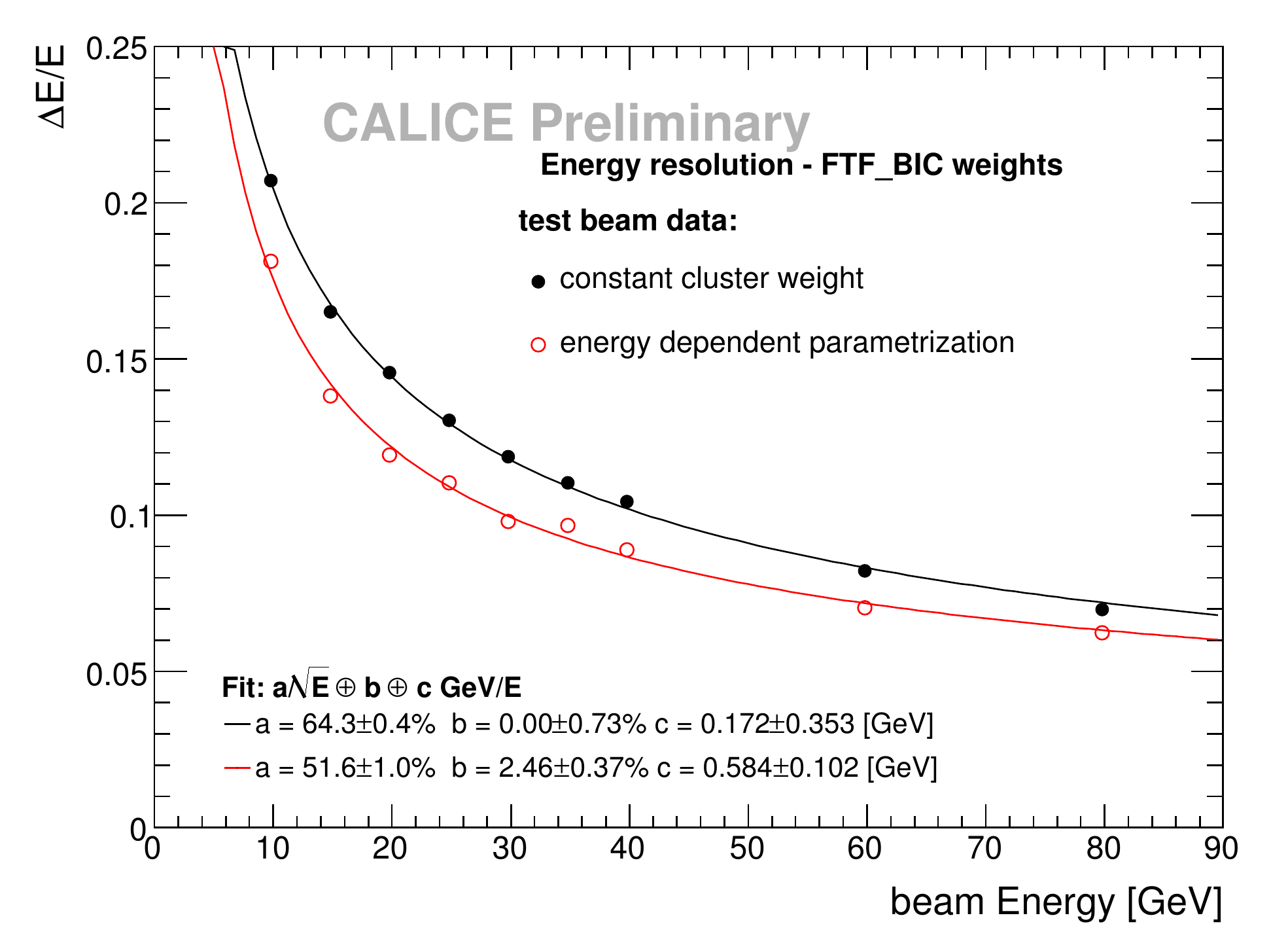}
  \caption{AHCAL energy resolution extracted from test-beam data with a single energy weight (filled circles) and energy weights based on the cluster energy density (open circles). }
  \label{fig_piRes}
\end{figure}

Weights for the visible energy deposited by hadrons are extracted from Monte Carlo simulations based on the cluster density and the cluster energy. Applying individual energy weights (depending on both the intially measured cluster energy and the cluster density) to each pion event in the detector yields a significant improvement of the pion linearity and energy resolution. Figure \ref{fig_piLin} shows the linearity before (filled circles) and after (open circles) applying this software compensation technique. The AHCAL response is normalised to the response to 15\,GeV pions. The weights are extracted from simulations using the FTF\_BIC physics list. Figure \ref{fig_piRes} shows the AHCAL energy resolution for single pions before (filled circles) and after (open circles) applying software compensation. The compensation algorithm reduces the stochastic term ($a/\sqrt{E}$) of the resolution by about 15\%. Further information on the software compensation technique presented in this paper is given in \cite{can21}. Studies of more sophisitcated software compensation algorithms are ongoing.

\section{Conclusions}
The CALICE collaboration has successfully built and operated an imaging calorimeter with analogue readout for hadrons. The pion data recorded with this detector allow for in-depth studies of the properties of hadronic cascades and the validation of Monte Carlo models with an unprecedented level of detail. With this highly granular calorimeter, it is possible to to determine the position of the first hard interaction, to investigate detailed shower profiles, and to measure the track multiplicity. The negative pion data presented in this paper cover the energy range from 8\,GeV to 80\,GeV. The analysis of pion data for lower energies down to 1\,GeV is ongoing.

The QGSP\_BERT physics list shows a strong effect from the transition to the LEP parameterisation in the energy region between 9\,GeV and 25\,GeV. The effects from model transitions  are reduced in the FTF-based physics lists (FTFP\_BERT and FTF\_BIC). For the observables presented in this paper, these physics lists yield the best overall performance with predictions agreeing with data within 5-10\%. The CHIPS physics list is promising since it has no transition between different models, but this model is still in a testing phase and requires further development.

The application of a simple software compensation algorithm to test-beam data improves the AHCAL energy resolution for single pions by 15\%.

\bibliographystyle{IEEEtran}
\bibliography{IEEEabrv,ANIMMA2011_proceedings_nfeege}

\begin{thebibliography}{10}
\providecommand{\url}[1]{#1}
\csname url@samestyle\endcsname
\providecommand{\newblock}{\relax}
\providecommand{\bibinfo}[2]{#2}
\providecommand{\BIBentrySTDinterwordspacing}{\spaceskip=0pt\relax}
\providecommand{\BIBentryALTinterwordstretchfactor}{4}
\providecommand{\BIBentryALTinterwordspacing}{\spaceskip=\fontdimen2\font plus
\BIBentryALTinterwordstretchfactor\fontdimen3\font minus
  \fontdimen4\font\relax}
\providecommand{\BIBforeignlanguage}[2]{{%
\expandafter\ifx\csname l@#1\endcsname\relax
\typeout{** WARNING: IEEEtran.bst: No hyphenation pattern has been}%
\typeout{** loaded for the language `#1'. Using the pattern for}%
\typeout{** the default language instead.}%
\else
\language=\csname l@#1\endcsname
\fi
#2}}
\providecommand{\BIBdecl}{\relax}
\BIBdecl

\bibitem{snowmass1}
J.-C. Brient and H.~Videau, ``{The calorimetry at the future e+ e- linear
  collider},'' in \emph{Proc. APS/DPF/DPB Summer Study on the Future of
  Particle Physics (Snowmass 2001)}, Colorado, USA, Jul. 2001.

\bibitem{pflow1}
V.~L. Morgunov, ``{Calorimetry design with energy-flow concept (imaging
  detector for high-energy physics)},'' in \emph{Proc. Tenth International
  Conference Pasadena}, California, USA, Mar. 2002.

\bibitem{pflow2}
M.~A. Thomson, ``{Particle Flow Calorimetry and the PandoraPFA Algorithm},''
  \emph{Nucl. Instrum. Meth.}, vol. A611, pp. 25--40, 2009.

\bibitem{sipm1}
G.~Bondarenko \emph{et~al.}, ``{Limited Geiger-mode microcell silicon
  photodiode: New results},'' \emph{Nucl. Instrum. Meth.}, vol. A422, pp.
  187--192, 2000.

\bibitem{ahcal1}
C.~Adloff \emph{et~al.}, ``{Construction and Commissioning of the CALICE Analog
  Hadron Calorimeter Prototype},'' \emph{JINST}, vol.~5, p. P05004, 2010.

\bibitem{ahcal2}
------, ``{Electromagnetic response of a highly granular hadronic
  calorimeter},'' \emph{JINST}, vol.~6, p. P04003, 2011.

\bibitem{can34}
\BIBentryALTinterwordspacing
N.~Feege, ``{Analysis of low-energetic electron and pion data collected with
  the AHCAL prototype at Fermilab},'' CALICE analysis note 34, Jun. 2011.
  [Online]. Available:
  \url{https://twiki.cern.ch/twiki/pub/CALICE/CaliceAnalysisNotes/CAN-034.pdf}
\BIBentrySTDinterwordspacing

\bibitem{mokka}
\BIBentryALTinterwordspacing
{Mokka - a GEANT4 application to simulate the full ILD geometry}. [Online].
  Available: \url{http://polzope.in2p3.fr:8081/MOKKA}
\BIBentrySTDinterwordspacing

\bibitem{geant4}
S.~Agostinelli \emph{et~al.}, ``{GEANT4: A simulation toolkit},'' \emph{Nucl.
  Instrum. Meth.}, vol. A506, pp. 250--303, 2003.

\bibitem{eudetmemo}
\BIBentryALTinterwordspacing
J.~Apostolakis \emph{et~al.}, ``{Validation of GEANT4 hadronic models using
  CALICE data},'' Eudet-Memo-2010-15. [Online]. Available:
  \url{http://www.eudet.org/e26/e28/e86887/e109012/EUDET-Memo-2010-15.pdf}
\BIBentrySTDinterwordspacing

\bibitem{can26}
\BIBentryALTinterwordspacing
A.~Kaplan, ``{Pion Showers in the CALICE AHCAL Prototype},'' CALICE analysis
  note 26, Nov. 2010. [Online]. Available:
  \url{https://twiki.cern.ch/twiki/pub/CALICE/CaliceAnalysisNotes/CAN-026.pdf}
\BIBentrySTDinterwordspacing

\bibitem{can22}
\BIBentryALTinterwordspacing
F.~Simon and L.~Weuste, ``{Identification of track segments in hadronic showers
  in the analog hadron calorimeter - algorithm and comparisons to
  simulations},'' CALICE analysis note 22, Jul. 2010. [Online]. Available:
  \url{https://twiki.cern.ch/twiki/pub/CALICE/CaliceAnalysisNotes/CAN-022.pdf}
\BIBentrySTDinterwordspacing

\bibitem{can21}
\BIBentryALTinterwordspacing
K.~Seidel and F.~Simon, ``{Software Compensation for Hadronic Showers in the
  CALICE AHCAL and Tail Catcher with Cluster-based Methods},'' CALICE analysis
  note 21, Apr. 2010. [Online]. Available:
  \url{https://twiki.cern.ch/twiki/pub/CALICE/CaliceAnalysisNotes/CAN-021.pdf}
\BIBentrySTDinterwordspacing

\end{thebibliography}

\end{document}